\documentclass[twocolumn,letter]{jpsj3}
\usepackage{graphicx}
\usepackage{dcolumn}
\usepackage{bm}
\usepackage{ulem}
\usepackage{braket}
\usepackage{amsmath}
\usepackage{color}

\title{
Cluster Toroidal Multipoles Formed by Electric-Quadrupole and Magnetic-Octupole Trimers: A Possible Scenario for Hidden Orders in Ca$_5$Ir$_3$O$_{12}$ 
}

\author{Satoru Hayami$^1$, Satoshi Tsutsui$^{2,3}$, Hiroki Hanate$^4$, Nobumoto Nagasawa$^2$, Yoshitaka Yoda$^2$, Kazuyuki Matsuhira$^4$}
\inst{
$^1$Graduate School of Science, Hokkaido University, Sapporo 060-0810, Japan \\
$^2$Japan Synchrotron Radiation Research Institute (JASRI), Sayo, Hyogo 679-5198, Japan \\
$^3$Institute of Quantum Beam Science, Graduate School of Science and Engineering, Ibaraki University, Hitachi, Ibaraki 316-8511, Japan \\
$^4$Graduate School of Engineering, Kyushu Institute of Technology, Kita-Kyushu, Fukuoka 804-8550, Japan
}

\abst{
Cluster multipole orderings composed of atomic high-rank multipole moments are theoretically investigated with a 5$d$-electron compound Ca$_5$Ir$_3$O$_{12}$ in mind. 
Ca$_5$Ir$_3$O$_{12}$ exhibits two hidden orders: One is an intermediate-temperature phase with time-reversal symmetry and the other is a low-temperature phase without time-reversal symmetry. 
By performing the symmetry and augmented multipole analyses for a $d$-orbital model under the hexagonal point group $D_{\rm 3h}$, we find that the 120$^{\circ}$-type ordering of the electric quadrupole corresponds to cluster electric toroidal dipole ordering with the electric ferroaxial moment, which can become the microscopic origin of the intermediate-temperature phase in Ca$_5$Ir$_3$O$_{12}$.  
Furthermore, based on ${}^{193}$Ir synchrotron-radiation-based M\"{o}ssbauer spectroscopy, we propose that the low-temperature phase in Ca$_5$Ir$_3$O$_{12}$ is regarded as a coexisting state with cluster electric toroidal dipole and cluster magnetic toroidal quadrupole, the latter of which is formed by the 120$^{\circ}$-type ordering of the magnetic octupole and accompanies a small uniform magnetization as a secondary effect. 
Our results provide a clue to two hidden phases in Ca$_5$Ir$_3$O$_{12}$. 
}

\begin{document}
\maketitle

A toroidal multipole moment, which is characterized by different spatial and time-reversal parities from conventional electric and magnetic multipole moments, has drawn considerable interest owing to its fascinating electronic structures, cross-correlation response, and transport~\cite{dubovik1975multipole,dubovik1986axial,dubovik1990toroid, Spaldin_0953-8984-20-43-434203,kopaev2009toroidal, Hlinka_PhysRevLett.113.165502,hayami2018microscopic}. 
There are two types of toroidal multipoles: magnetic toroidal (MT) multipoles with the same symmetry property as time-reversal-odd polar tensors and electric toroidal (ET) multipoles with the same symmetry as time-reversal-even axial tensors. 
Among them, the odd-parity toroidal multipoles, i.e., the odd-rank MT and even-rank ET multipoles, have been extensively studied, since they give rise to physical phenomena related to the spatial-parity breaking, such as the magnetoelectric effect under the MT dipole~\cite{Fiebig0022-3727-38-8-R01, Eerenstein2006multiferroic, tokura2014multiferroics, Yanase_JPSJ.83.014703, Hayami_PhysRevB.90.024432, Hayami_doi:10.7566/JPSJ.84.064717,thole2018magnetoelectric,yatsushiro2019atomic,thole2020concepts, Hayami_PhysRevB.105.104428}, nonlinear (spin) transport~\cite{Wang_PhysRevLett.127.277201, Liu_PhysRevLett.127.277202, Yatsushiro_PhysRevB.105.155157, Hayami_PhysRevB.106.024405, Hayami_PhysRevB.106.014420, Kondo_PhysRevResearch.4.013186, Hayami_doi:10.7566/JPSJ.91.094704, kirikoshi2022microscopic} and nonreciprocal magnon excitations under the MT dipole and octupole~\cite{Miyahara_JPSJ.81.023712, Miyahara_PhysRevB.89.195145, Hayami_doi:10.7566/JPSJ.85.053705, Matsumoto_PhysRevB.101.224419, Matsumoto_PhysRevB.104.134420, Hayami_PhysRevB.105.014404}, and Edelstein effect and rotation-field induced electric polarization under the ET monopole and quadrupole~\cite{yoda2015current, furukawa2017observation, hitomi2019magnetoelectric, Ishitobi_doi:10.7566/JPSJ.88.063708, yatsushiro2020odd, Oiwa_PhysRevLett.129.116401}. 
These unconventional electronic orderings have been proposed and identified in metallic materials, such as the MT dipole orderings in UNi$_4$B~\cite{Mentink1994, Oyamada2007, Hayami_PhysRevB.90.024432, saito2018evidence, Yanagisawa_PhysRevLett.126.157201, tabata2021x, ota2022zero, ishitobi2022triple}, Ce$_3$TiBi$_5$~\cite{motoyama2018magnetic, shinozaki2020magnetoelectric, shinozaki2020study, Hayami_doi:10.7566/JPSJ.91.123701}, and CeCoSi~\cite{tanida2018substitution, yatsushiro2020odd, nikitin2020gradual, Manago_doi:10.7566/JPSJ.90.023702,  Yatsushiro_PhysRevB.102.195147, Matsumura_doi:10.7566/JPSJ.91.064704, Tanida_doi:10.7566/JPSJ.91.094701, Yatsushiro_doi:10.7566/JPSJ.91.104701} and ET quadrupole orderings in Cd$_2$Re$_2$O$_7$~\cite{Yamaura_PhysRevLett.108.247205, Yamaura_PhysRevB.95.020102, hiroi2017pyrochlore, harter2017parity, Matteo_PhysRevB.96.115156, matsubayashi2018split, Hayami_PhysRevLett.122.147602, Hirose_PhysRevB.105.035116, hirai2022successive}. 

Meanwhile, the even-parity toroidal multipoles, i.e., the even-rank MT and odd-rank ET multipoles, have also been recognized to exhibit various off-diagonal responses and transports in spite of the presence of spatial inversion symmetry. 
For example, the MT quadrupole (MTQ) gives rise to a symmetric spin splitting in the electronic band structure to generate the spin current even without relativistic spin--orbit coupling (SOC)~\cite{naka2019spin, hayami2019momentum, Hayami2020b, hayami2022spinconductivity}. 
In addition, the ET dipole (ETD) and octupole induce unconventional transverse responses~\cite{cheong2021permutable, Hayami_doi:10.7566/JPSJ.91.113702, Hayami_PhysRevB.106.144402}, such as the antisymmetric thermopolarization~\cite{Nasu_PhysRevB.105.245125}. 
However, candidate materials with even-parity toroidal multipoles have been rare compared to those with odd-parity ones; CaMn$_7$O$_{12}$~\cite{Johnson_PhysRevLett.108.067201}, RbFe(MoO$_4$)$_2$~\cite{jin2020observation,Hayashida_PhysRevMaterials.5.124409}, and NiTiO$_3$~\cite{hayashida2020visualization, Hayashida_PhysRevMaterials.5.124409, yokota2022three} are a few examples to possess the ETD in bulk and no materials with the MTQ have been recognized~\cite{hayami2022spinconductivity}. 
Thus, searching for compounds with even-parity toroidal orderings is highly desired.  

Ca$_5$Ir$_3$O$_{12}$ has a hexagonal structure ($P\bar{6}2m$), which might be a prototype to possess both ETD and MTQ, as will be discussed. 
This material undergoes two phase transitions at $105$~K and $7.8$~K~\cite{wakeshima2003electrical, Matsuhira_doi:10.7566/JPSJ.87.013703}: The former transition was identified as the onset of the ETD ordering with the $\sqrt{3}a\times \sqrt{3}a\times 3c$ superstructure in the hexagonal coordinate~\cite{Hasegawa_doi:10.7566/JPSJ.89.054602, Hanate_doi:10.7566/JPSJ.89.053601, hanate2021first}, although relevant electronic degrees of freedom for the ETD ordering have been still lacking.  
The latter was identified as the antiferromagnetic ordering~\cite{Cao_PhysRevB.75.134402, Franke_PhysRevB.83.094416, HANATE2022170072}, where a detailed structure has not been revealed; no magnetic reflections have been observed by powder neutron diffraction experiments~\cite{comment_Wakeshima} and a possibility of multipole magnetic ordering was implied~\cite{HANATE2022170072}, which is similar to the hidden order in URu$_2$Si$_2$~\cite{Palstra_PhysRevLett.55.2727,mydosh2014hidden}. 
The emergence of such peculiar orderings might be attributed to a comparable energy scale among the SOC, exchange interaction, and transfer arising from the Ir $5d$ electrons, as demonstrated by the density functional theory calculations~\cite{Charlebois_PhysRevB.104.075153}. 

In the present study, we investigate the possibility of even-parity ETD and MTQ orderings in Ca$_5$Ir$_3$O$_{12}$. 
Based on symmetry analysis, cluster multipole theory, and experimental observations including ${}^{193}$Ir synchrotron-radiation-based M\"{o}ssbauer spectroscopy (SRMS), we propose that the intermediate-temperature (IT) phase corresponds to a cluster ETD and the low-temperature (LT) phase corresponds to a cluster MTQ; the former is characterized by an electric-quadrupole (EQ) trimer, while the latter is by a magnetic-octupole (MO) trimer. 
We show that the system under both the ETD and MTQ exhibits a spontaneous magnetization along the $z$ direction. 
We also discuss possible superlattice structures for two ordered phases in Ca$_5$Ir$_3$O$_{12}$.

\begin{table}[htb!]
\caption{
Irreducible representation (IR) of the atomic multipoles (MPs) in the basis of $\{\ket{d_{zx} \uparrow}, \ket{d_{yz} \uparrow}, \ket{d_{zx} \downarrow}, \ket{d_{yz} \downarrow}\}$ under the point groups $D_{\rm 3h}$ and $C_{\rm 3h}$. 
The superscript for the IR represents the time-reversal parity. 
The second column represents the rank of the multipole. 
The third column represents the basis wave functions, where $\bm{R}=(R_x, R_y, R_z)$ corresponds to a time-reversal-odd axial vector. 
The fourth column represents the representation of the Pauli matrices $\tau_\mu \sigma_{\nu}$ for $\mu, \nu=0,x,y,z$; $\rho_{\pm}=(\tau_z\sigma_x\pm \tau_x \sigma_y)/2$ and $\rho'_{\pm}=(\pm \tau_x\sigma_x+ \tau_z \sigma_y)/2$. 
}
\label{tab: mp}
\centering
\renewcommand{\arraystretch}{1.0}
 \begin{tabular}{cccccllllll}
 \hline  \hline
MP & rank & basis & $\tau_\mu \sigma_\nu$ & $D_{\rm 3h}$ ($C_{\rm 3h}$) \\ \hline 
$Q_0$ & 0 & 1 &  $\tau_0\sigma_0$ & ${\rm A}'^+_{1}$ (${\rm A}'^+$) \\ 
$Q_{v}$ & 2 & $\frac{1}{2}(x^2-y^2)$ &  $\tau_z\sigma_0$ & ${\rm E}'^+$ (${\rm E}'^+$) \\ 
$Q_{xy}$ & 2 & $xy$ &  $\tau_x\sigma_0$ & ${\rm E}'^+$ (${\rm E}'^+$) \\ 
$M_z$ & 1 & $R_z$   & $\tau_y\sigma_0$ & ${\rm A}'^-_{2}$ (${\rm A}'^-$) \\ 
$M'_x$& 1 & $R_x$ &  $\tau_0\sigma_x$ & ${\rm E}''^-$ (${\rm E}''^-$) \\ 
$M'_y$& 1 & $R_y$  &  $\tau_0\sigma_y$ & ${\rm E}''^-$ (${\rm E}''^-$) \\ 
$M'_z$& 1 & $R_z$ &  $\tau_0\sigma_z$ & ${\rm A}'^-_{2}$ (${\rm A}'^-$)\\ 
$Q'_0$& 0 & $1$ &  $\tau_y\sigma_z$ & ${\rm A}'^+_{1}$ (${\rm A}'^+$)\\ 
$Q'_{zx}$& 2 & $zx$ &  $\tau_y\sigma_x$ & ${\rm E}''^+$ (${\rm E}''^+$) \\ 
$Q'_{yz}$& 2 & $yz$ &  $\tau_y\sigma_y$ & ${\rm E}''^+$ (${\rm E}''^+$) \\ 
$M'_{xyz}$& 3 & $R_x R_y R_z$ &  $\tau_x\sigma_z$ &  ${\rm E}'^-$ (${\rm E}'^-$) \\ 
$M'^{\beta}_z$& 3 & $\frac{1}{2}(R^2_x-R_y^2)R_z $ &  $\tau_z\sigma_z$ &  ${\rm E}'^-$ (${\rm E}'^-$) \\ 
$M'_{3a}$& 3 & $R_x(R_x^2-3R_y^2)$ &  $\rho_-$ & ${\rm A}''^-_{1}$ (${\rm A}''^-$) \\ 
$M'_{3b}$& 3 & $R_y(3R_x^2-R_y^2)$ &  $\rho'_+$ & ${\rm A}''^-_{2}$ (${\rm A}''^-$) \\ 
$M'_{3v}$& 3 & $R_y(5z^2-r^2)$ &  $\rho'_-$ & ${\rm E}''^-$ (${\rm E}''^-$) \\ 
$M'_{3u}$& 3 & $R_x(5z^2-r^2)$ &  $\rho_+$ & ${\rm E}''^-$ (${\rm E}''^-$) \\ 
 \hline
\hline 
\end{tabular}
\end{table}

Let us start by considering a three-sublattice triangular cluster with five $d$ orbitals, $(d_u, d_v, d_{yz}, d_{zx}, d_{xy})$ for $u=3z^2-r^2$ and $v=x^2-y^2$, under the point group $D_{3\rm h}$. 
We denote the sublattice index as $i=1$--$3$. 
In the spinless Hilbert space, the irreducible representations (IRs) of the five $d$ orbitals are $d_u \in {\rm A}'_1$, $(d_v, d_{xy}) \in {\rm E}'$, and $(d_{yz}, d_{zx}) \in {\rm E}''$.  
Then, active atomic multipoles with time-reversal even are rank-0 electric monopole, rank-2 EQ, and rank-4 electric hexadecapole, which belong to the IR ${\rm A}'^+_1$, ${\rm A}''^+_1$, ${\rm A}''^+_2$, ${\rm E}'^+$, and ${\rm E}''^+$ according to the type of multipoles; the superscript $\pm$ represents the time-reversal parity~\cite{Hayami_PhysRevB.98.165110, Yatsushiro_PhysRevB.104.054412}. 
Among them, the atomic multipoles belonging to the ${\rm E}'^+$ representation can construct the ${\rm A}'^+_2$ representation corresponding to the ETD within a triangular cluster. 
Thus, the Hilbert space should include such multipole degrees of freedom belonging to the ${\rm E}'^+$ representation to describe the ETD ordering observed in Ca$_5$Ir$_3$O$_{12}$~\cite{hanate2021first}, where each triangle unit of the Ir ions is supposed to have the ETD based on the Raman scattering experiments~\cite{Hasegawa_doi:10.7566/JPSJ.89.054602}.  

Based on this symmetry analysis, we consider an effective model to include the multipoles under the ${\rm E}'^+$ representation. 
Specifically, we adopt a minimal two-orbital basis consisting of $d_{zx}$ and $d_{yz}$, which forms $\ket{\pm 1 \sigma}=\left(\mp \ket{d_{zx} \sigma} - i\ket{d_{yz}\sigma}\right)/\sqrt{2}$ for $\sigma=\uparrow, \downarrow$; $\pm 1$ represents the magnetic quantum number~\cite{Hayami_PhysRevB.90.081115}.
For this atomic basis, there are 4 spinless and 12 spinful multipole degrees of freedom. 
By introducing the Pauli matrices $\tau_\mu$ and $\sigma_{\nu}$ for $\mu,\nu=0,x,y,z$ in the orbital and spin space for the basis $\{\ket{d_{zx} \uparrow}, \ket{d_{yz} \uparrow}, \ket{d_{zx} \downarrow}, \ket{d_{yz} \downarrow}\}$, respectively ($\tau_0$ and $\sigma_0$ denote the $2\times 2$ identity matrices), each multipole is described by a linear combination of $\tau_\mu \sigma_\nu$, as shown in Table~\ref{tab: mp}~\cite{hayami2016emergent,kusunose2020complete}. 
The prime symbol in the multipole in Table~\ref{tab: mp} means the spinful multipoles.
The multipoles belonging to the ${\rm E}'^+$ representation are two EQs: $Q_{v}$ and $Q_{xy}$~\cite{comment_EH}.

\begin{figure}[t!]
\begin{center}
\includegraphics[width=1.0 \hsize ]{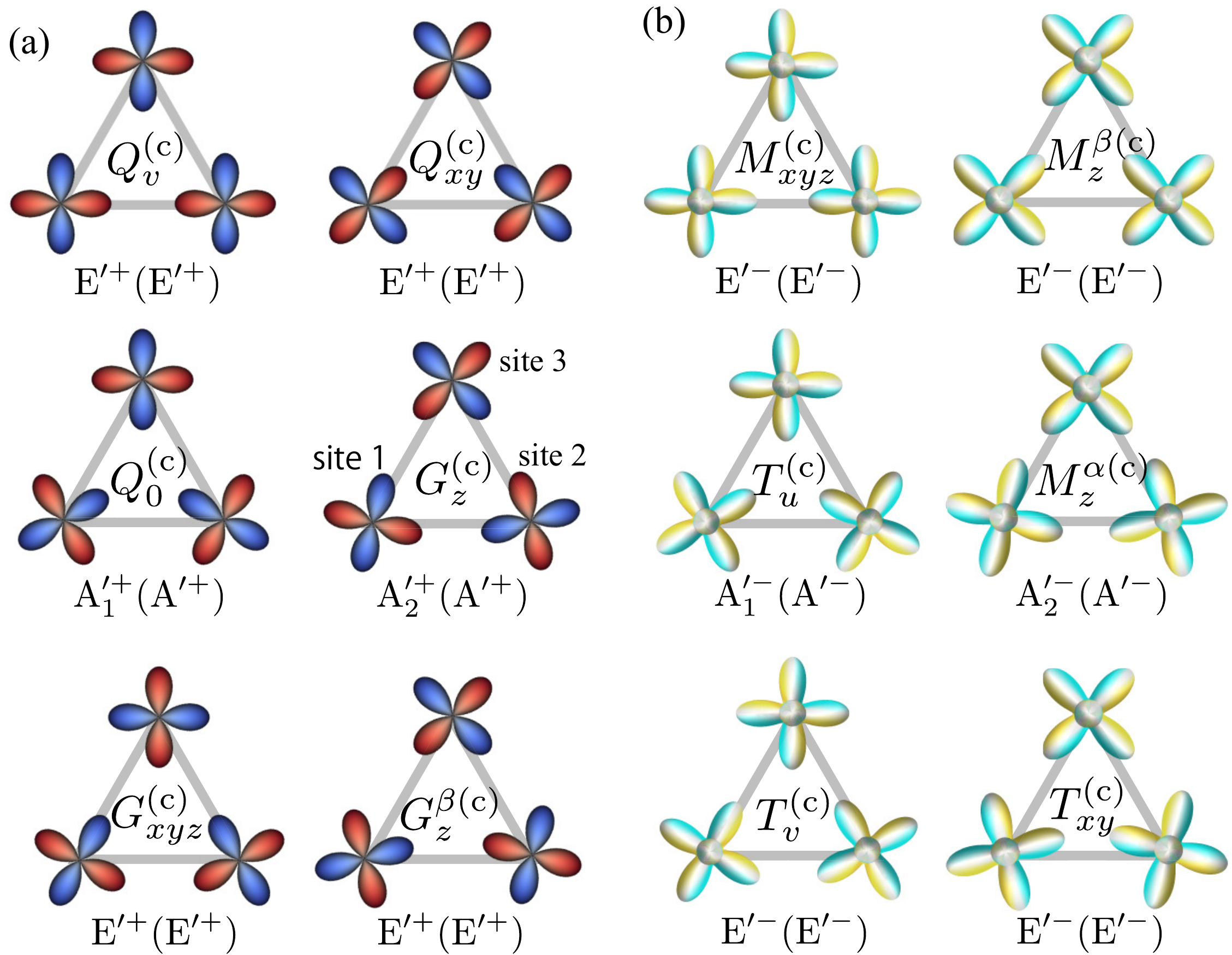} 
\caption{
\label{fig: cluster}
(Color online) 
Cluster multipoles constructed from (a) the EQs $(Q_{v}, Q_{xy})$ and (b) MOs $(M'_{xyz}, M'^{\beta}_z)$. 
The colors for $(Q_{v}, Q_{xy})$ and $(M'_{xyz}, M'^{\beta}_z)$ represent the phase factor of the wave function and $z$ component of magnetic dipole moments, respectively. 
The IRs in each cluster multipole under the point groups $D_{\rm 3h}$ and $C_{\rm 3h}$ are also shown. 
}
\end{center}
\end{figure}

Next, let us consider the three-sublattice structure of $Q_{v}$ and $Q_{xy}$.  
We show possible three-sublattice structures in Fig.~\ref{fig: cluster}(a) based on cluster multipole theory~\cite{Suzuki_PhysRevB.95.094406, Suzuki_PhysRevB.99.174407}, where the superscript (c) of the multipole represents the cluster multipole. 
When superposing $(Q_{v}, Q_{xy})=(\sqrt{3}/2, -1/2)$ for site $1$, $(-\sqrt{3}/2, -1/2)$ for site 2, and $(0,1)$ for site 3, we obtain the cluster ETD $G^{\rm (c)}_z$ belonging to the ${\rm A}'^+_{2}$ representation, which reduces the symmetry to $C_{\rm 3h}$. 
Since the other three-sublattice structures are incompatible with the ${\rm A}'^+_{2}$ representation, the orbital trimer consisting of the EQs is a natural candidate of the IT phase in Ca$_5$Ir$_3$O$_{12}$.

\begin{figure}[t!]
\begin{center}
\includegraphics[width=0.7 \hsize ]{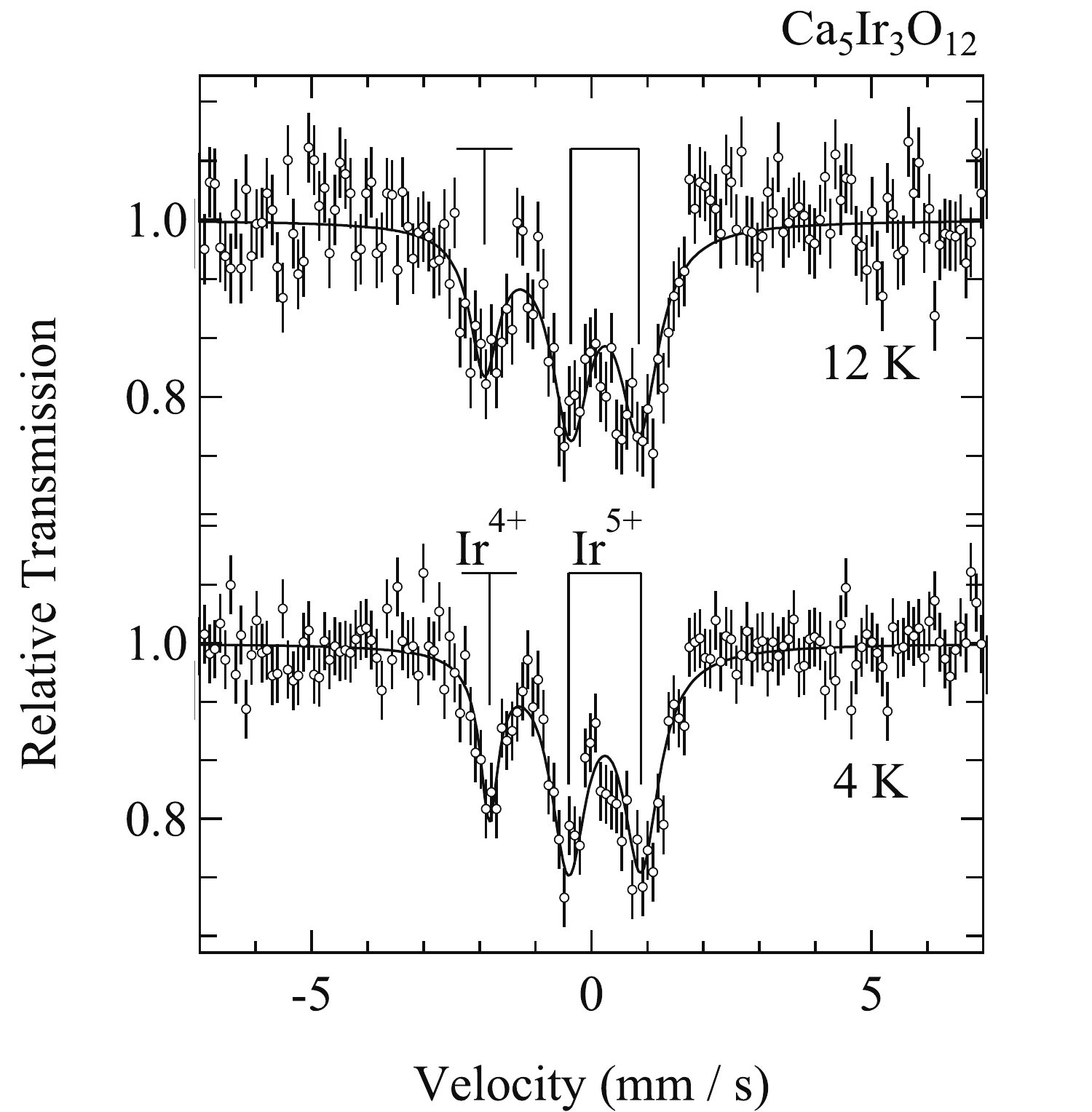} 
\caption{
\label{fig: SR_MS}
${}^{193}$Ir SR-based M\"{o}ssbauer spectra of Ca$_5$Ir$_3$O$_{12}$ at 4 and 12~K. 
Circles depict measured data. 
Solid lines depict fitting curves.
}
\end{center}
\end{figure}

The Kramers degeneracy remains in the $G^{\rm (c)}_{z}$ order owing to the presence of time-reversal symmetry. 
Then, the further phase transition breaking time-reversal symmetry is expected, which corresponds to the LT phase in Ca$_5$Ir$_3$O$_{12}$. 
To narrow down possible order parameters, we perform ${}^{193}$Ir SRMS at BL35XU of SPring-8. 
The experimental details were reported in Ref.~\citen{tsutsui2021application}~\cite{  comment_spring8_2}. 
Figure~\ref{fig: SR_MS} shows ${}^{193}$Ir SRMS in the IT phase at 12~K and the LT phase at $4$~K measured with 4 bunch $\times$ 84 train mode in SPring-8~\cite{comment_spring8}, where the data for $T \geq 30$ K were presented in Ref.~\citen{Tsutsui_doi:10.7566/JPSJ.90.083701}. 
The spectra do not qualitatively change in the phase transition, which demonstrates that hyperfine magnetic field is absent at ${}^{193}$Ir below the magnetic transition temperature of 7.8 K. 
The resolution limit of the magnetic moment for Ir$^{4+}$ is evaluated around 0.01~$\mu_{\rm B}$ by using its hyperfine coupling constant as 85~T/$\mu_{\rm B}$~\cite{wagner1983mossbauer, Ye_PhysRevB.85.180403, Ye_PhysRevB.87.140406} and the magnetic splitting 1~mm/s in ${}^{193}$Ir SRMS spectra corresponding to 0.85~T, which is comparable to the magnitude of the magnetic moments observed by previous $\mu$SR studies with the dipole approximation~\cite{Franke_PhysRevB.83.094416}.
Since such a situation is similar to NpO$_2$~\cite{dunlap1968hyperfine, Friedt_PhysRevB.32.257, kopmann1998magnetic}, the higher-rank magnetic multipole ordering is expected.

In addition, we take into account two conditions: (i) tiny uniform magnetization~\cite{comment_Amitsuka} and (ii) threefold symmetric cluster structure~\cite{Hasegawa_doi:10.7566/JPSJ.89.054602}. 
Based on these conditions, it is natural to consider that the primary magnetic order parameter is described by the three-sublattice ordering of MOs rather than the magnetic dipoles. 
Among the MOs $(M'_{xyz}, M'^{\beta}_z, M'_{3a}, M'_{3b}, M'_{3u}, M'_{3v})$ in Table~\ref{tab: mp}, only $M'_{xyz}$ and $M'^{\beta}_z$ belonging to ${\rm E}'^-$ satisfy the above conditions. 
We present six three-sublattice structures of $M'_{xyz}$ and $M'^{\beta}_z$ in Fig.~\ref{fig: cluster}(b), where the cluster MTQ $T^{\rm (c)}_{u}$ and the cluster MO $M^{\alpha{\rm (c)}}_{z}$ are allowed to have a uniform magnetization along the $z$ direction ($M_z$ and $M'_z$); $T^{\rm (c)}_{u}$ ($M^{\alpha{\rm (c)}}_{z}$) is constituted of $(M'^{\beta}_{z}, M'_{xyz})=(\sqrt{3}/2, -1/2)$ [$(-1/2,-\sqrt{3}/2)$] for site $1$, $(-\sqrt{3}/2, -1/2)$ [$(-1/2,\sqrt{3}/2)$] for site 2, and $(0,1)$ [$(1,0)$] for site 3. 
In this case, $M_z$ and $M'_z$ correspond to the secondary order parameters. 
Although the resultant magnetic point group is the same between $T^{\rm (c)}_{u}$ and $M^{\alpha{\rm (c)}}_{z}$, their origin of spontaneous magnetization is different from each other, as detailed below. 

We model the above situation in the triangular cluster for the basis $\{\ket{d_{zx} \uparrow}, \ket{d_{yz} \uparrow}, \ket{d_{zx} \downarrow}, \ket{d_{yz} \downarrow}\}$. 
The minimum Hamiltonian is represented as 
\begin{align}
\label{eq: Ham}
\mathcal{H}&= \mathcal{H}_{\rm SOC} + \mathcal{H}_t + \mathcal{H}^{\rm Q}_{\rm MF} + \mathcal{H}^{\rm M}_{\rm MF}, \\
\mathcal{H}_{\rm SOC} &= \lambda \sum_i  l^z_i s^z_i,\\
\mathcal{H}_t &=\sum_{ij\alpha\beta \sigma}  \left(t^{\alpha\beta}_{ij}c^\dagger_{i\alpha \sigma}c^{}_{j\beta\sigma}+ {\rm h.c.}\right), \\
\label{eq: Ham_Q}
\mathcal{H}^{\rm Q}_{\rm MF}&= -h_{\rm Q}\sum_{i\alpha \beta \sigma} (\bm{e}_i \cdot \bm{\tilde{\tau}})^{\alpha\beta} c^{\dagger}_{i\alpha \sigma} c^{}_{i \beta \sigma}, \\
\label{eq: Ham_M}
\mathcal{H}^{\rm M}_{\rm MF}&= -h_{\rm M} \sum_{i\alpha \beta \sigma} (\bm{e}_i\cdot \bm{\tilde{\tau}}  )^{\alpha\beta} p(\sigma) c^{\dagger}_{i\alpha \sigma} c^{}_{i \beta \sigma} \nonumber \\
&\ \ \ -h'_{\rm M} \sum_{i\alpha \beta \sigma} (\bm{e}_i \times \bm{\tilde{\tau}} )_z^{\alpha\beta} p(\sigma) c^{\dagger}_{i\alpha \sigma} c^{}_{i \beta \sigma},
\end{align}
where $c^\dagger_{i\alpha\sigma}$ and $c_{i\alpha\sigma}$ are creation and annihilation operators of electrons at site $i=$1--$3$, orbital $\alpha=v, xy$, and spin $\sigma=\uparrow, \downarrow$. 
The total Hamiltonian $\mathcal{H}$ consists of four terms. 
The first term $\mathcal{H}_{\rm SOC}$ in Eq.~(\ref{eq: Ham}) represents the SOC, where $\bm{l}_i=(l_i^x, l_i^y, l_i^z)$ and $\bm{s}_i=(s_i^x, s_i^y, s_i^z)$ are orbital and spin angular momentum operators, respectively. 
The second term $\mathcal{H}_t$ in Eq.~(\ref{eq: Ham}) represents the hopping to satisfy the Slater-Koster parameters: $t_{dd\pi}=1$. 
The third term $\mathcal{H}^{\rm Q}_{\rm MF}$ in Eq.~(\ref{eq: Ham}) stands for the mean field corresponding to $G_z^{\rm (c)}$, where we introduce $\bm{\tilde{\tau}}=(-\tau_z, \tau_x)$ and the unit vectors for $i$th site $\bm{e}_i$; $\bm{e}_1=(-\sqrt{3}/2,-1/2)$, $\bm{e}_2=(\sqrt{3}/2, -1/2)$, and $\bm{e}_3=(0,1)$. 
The fourth term $\mathcal{H}^{\rm M}_{\rm MF}$ in Eq.~(\ref{eq: Ham}) stands for the mean fields corresponding to $T_u^{\rm (c)}$ in the first term and $M_z^{\alpha{\rm (c)}}$ in the second term; $p(\sigma)=+ 1 $ $(-1)$ for $\sigma=\uparrow$ ($\downarrow$).  
It is noted that $h_{\rm Q}\neq 0$ in the IT phase, while $h_{\rm M}\neq 0$ and/or $h'_{\rm M}\neq 0$ in addition to $h_{\rm Q}\neq 0$ in the LT phase. 
The amplitudes of the mean fields are set as $h_{\rm Q}$ in Eq.~(\ref{eq: Ham_Q}) and $h^{(')}_{\rm M}$ in Eq.~(\ref{eq: Ham_M}). 

\begin{figure}[t!]
\begin{center}
\includegraphics[width=1.0 \hsize ]{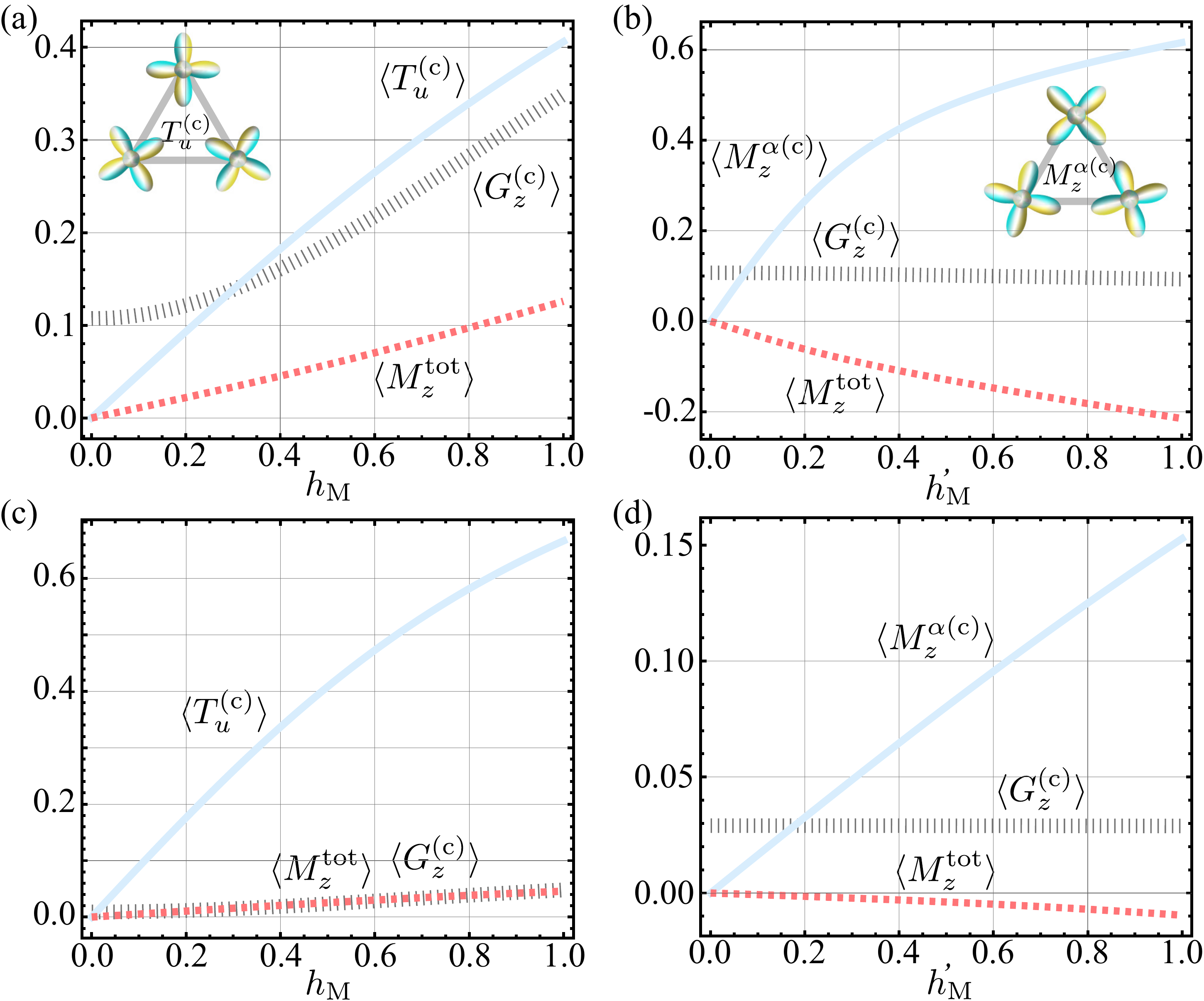} 
\caption{
\label{fig: mag}
(Color online) 
(a,c) $h_{\rm M}$ dependence of $\langle G_z^{\rm (c)} \rangle$, $\langle T_u^{\rm (c)} \rangle$, and $\langle M^{\rm tot}_z \rangle$ in the $T_u^{\rm (c)}$ order for $h'_{\rm M}=0$ at (a) $\lambda=5$ and $h_{\rm Q}=0.3$ and (c) $\lambda=1$ and $h_{\rm Q}=0.01$. 
(b,d) $h'_{\rm M}$ dependence of $\langle G_z^{\rm (c)} \rangle$, $\langle M_z^{\alpha{\rm (c)}} \rangle$, and $\langle M^{\rm tot}_z \rangle$ in the $M^{\alpha{\rm (c)}}_z$ order for $h_{\rm M}=0$ and $h_{\rm Q}=0.3$ at (b) $\lambda=5$ and (d) $\lambda=20$. 
}
\end{center}
\end{figure}

For the model in Eq.~(\ref{eq: Ham}), we investigate the behavior of the uniform magnetization $\langle M^{\rm tot}_z\rangle=\langle l^z + 2 s^z \rangle= \langle \sum_i (l^z_i + 2 s^z_i) \rangle$ with $\langle l^z_1 + 2 s^z_1 \rangle=\langle l^z_2 + 2 s^z_2 \rangle=\langle l^z_3 + 2 s^z_3 \rangle$ in the LT phase by setting the temperature $T=0.1$, where $\langle M^{\rm tot}_z \rangle = {\rm Tr}[e^{-E_k/T} M^{\rm tot}_z]/Z$ ($Z$ is the partition function, $T$ is the temperature, and $E_k$ is the $k$th eigenenergies). 
Figure~\ref{fig: mag}(a) shows the $h_{\rm M}$ dependence of $\langle M^{\rm tot}_z \rangle$ as well as $\langle G_z^{\rm (c)} \rangle$ and $\langle T_u^{\rm (c)} \rangle$ in the $T_u^{\rm (c)}$ order at $h'_{\rm M}=0$, $\lambda=5$, and $h_{\rm Q}=0.3$; $\langle G_z^{\rm (c)} \rangle$ and $\langle T_u^{\rm (c)} \rangle$ correspond to the 120$^{\circ}$ distributions of the atomic $(Q_v, Q_{xy})$ and $(M'_{xyz}, M'^{\beta}_z)$, respectively, as shown in Fig.~\ref{fig: cluster}. 
The result clearly indicates that $\langle M^{\rm tot}_z \rangle$ increases while $\langle T_u^{\rm (c)} \rangle$ is developed. 
The similar result is also obtained in the $M^{\alpha{\rm (c)}}_z$ order by setting $h_{\rm M}=0$ but $h'_{\rm M}\neq 0$; $\langle M^{\rm tot}_z \rangle$ becomes nonzero for $\langle M^{\alpha{\rm (c)}}_z \rangle \neq 0$, as shown in Fig.~\ref{fig: mag}(b).

The different behavior of $\langle M^{\rm tot}_{z} \rangle$ between the $T_u^{\rm (c)}$ and $M^{\alpha{\rm (c)}}_z$ orders is found when $\langle G_z^{\rm (c)} \rangle \to -\langle G_z^{\rm (c)} \rangle$; $\langle M^{\rm tot}_{z} \rangle \to -\langle M^{\rm tot}_{z}\rangle$ in the $T_u^{\rm (c)}$ order, whereas $\langle M^{\rm tot}_{z} \rangle \to \langle M^{\rm tot}_{z}  \rangle$ in the $M^{\alpha{\rm (c)}}_z$ order. 
In other words, the uniform magnetization is canceled out for different domains of $\langle G_z^{\rm (c)} \rangle$ in the $T_u^{\rm (c)}$ order, while it is not in the $M^{\alpha{\rm (c)}}_z$ order. 

The magnitude of $\langle M^{\rm tot}_z \rangle$ is dependent on the model parameters $t_{dd\pi}$, $\lambda$, and $h_{\rm Q}$. 
We present the cases of small $\langle M^{\rm tot}_z \rangle$ by taking $h_{\rm Q}=0.01$ and $\lambda=1$ in the $T_u^{\rm (c)}$ order in Fig.~\ref{fig: mag}(c) and taking $\lambda=20$ in the $M^{\alpha{\rm (c)}}_z$ order in Fig.~\ref{fig: mag}(d) as examples. 
In both ordered cases, the crystal Hall effect is expected even for negligibly small $\langle M^{\rm tot}_z \rangle$, which might be a signature of the $T_u^{\rm (c)}$ and/or $M^{\alpha{\rm (c)}}_z$ orders~\cite{nakatsuji2015large, Naka_PhysRevB.102.075112, vsmejkal2020crystal, Hayami_PhysRevB.103.L180407}. 

Let us discuss the model parameter dependence of $\langle M^{\rm tot}_z \rangle$ for both ordered states.  
By analytically expanding the trace, one can obtain the essential model parameters for inducing $\langle M^{\rm tot}_z \rangle$~\cite{Hayami_PhysRevB.101.220403, Hayami_PhysRevB.102.144441, Oiwa_doi:10.7566/JPSJ.91.014701}. 
In the $T_u^{\rm (c)}$ order, we find that $\langle s^z \rangle$ ($\langle l^z \rangle$) is proportional to $h_{\rm Q} h_{\rm M}$ ($\lambda h_{\rm Q} h_{\rm M}$). 
Thus, nonzero $\langle G_z^{\rm (c)} \rangle$ is essentially important but the hopping is not for inducing $\langle M^{\rm tot}_z \rangle $. 
This is understood from the atomic limit by taking $t_{dd\pi} \to 0$. 
When considering site 3, one finds that a superposition of $\tau_x \sigma_0$ corresponding to $Q_{xy}$ and $\tau_x \sigma_z$ corresponding to $M'_{xyz}$ naturally leads to $\tau_0 \sigma_z$ in their coupling. 
This is also consistent with the symmetry argument; $\langle M^{\rm tot}_z \rangle $ must vanish for $\langle G_z^{\rm (c)} \rangle=0$, since the representation of $T_u^{\rm (c)}$ is different from that of $M^{\rm tot}_z$. 
This is why the sign of $\langle M^{\rm tot}_z \rangle $ depends on that of $\langle G_z^{\rm (c)} \rangle$. 

On the other hand, the different essential model parameters for $\langle M^{\rm tot}_z \rangle$ are obtained in the $M_z^{\alpha{\rm (c)}}$ order; $\langle s^z \rangle$ ($\langle l^z \rangle$) is proportional to $h'_{\rm M}t_{dd\pi}^2$ ($\lambda h'_{\rm M} t_{dd\pi}^2$). 
Thus, the precursor $\langle G_z^{\rm (c)} \rangle$ is not necessarily for nonzero $\langle M^{\rm tot}_z \rangle$ in this case. 
Moreover, one finds that the cluster structure connected by the hopping is important in this case; $\langle M^{\rm tot}_z \rangle$ vanishes for $t_{dd\pi} = 0$.

\begin{figure}[t!]
\begin{center}
\includegraphics[width=0.7 \hsize ]{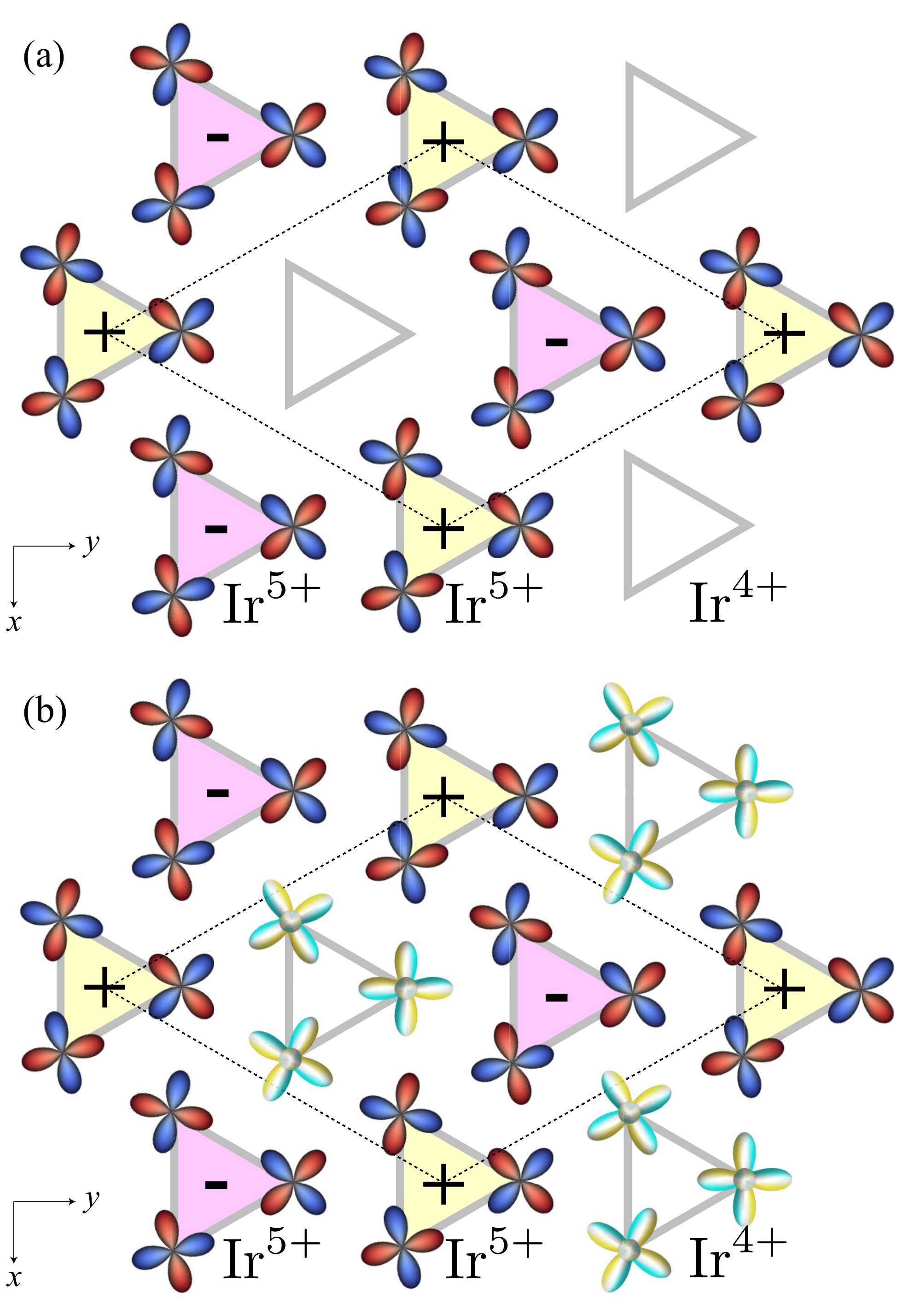} 
\caption{
\label{fig: Ca5Ir3O12}
(Color online) 
Schematic two-dimensional pictures of (a) a possible ordering with $G_z^{\rm (c)}$ in the IT phase and (b) that with $G_z^{\rm (c)}$ and $T_u^{\rm (c)}$ in the LT phase in Ca$_5$Ir$_3$O$_{12}$. 
The sign denoted in the triangular plaquette represents the sign of $G_z^{\rm (c)}$. 
The dashed rhombus represents the unit cell.
}
\end{center}
\end{figure}

We apply the above results to the crystal system with Ca$_5$Ir$_3$O$_{12}$ in mind. 
In the present scenario, the IT phase in Ca$_5$Ir$_3$O$_{12}$ is described by the density waves of the cluster ETD (EQ trimer) $\langle G_z^{\rm (c)} \rangle$ at $\bm{q}=(1/3,1/3,1/3)$, as shown in Fig.~\ref{fig: Ca5Ir3O12}(a). 
The periodic alignment of $(+\langle G_z^{\rm (c)} \rangle, -\langle G_z^{\rm (c)} \rangle, 0)$ leads to a charge disproportionation of the Ir ion to satisfy ${\rm Ir}^{4+}:{\rm Ir}^{5+}=1:2$, which is consistent with SRMS~\cite{Tsutsui_doi:10.7566/JPSJ.90.083701}. 
When the temperature is lowered, the remaining triangle occupied by ${\rm Ir}^{4+}$ is expected to show the $T^{\rm (c)}_{u}$ (or $M_z^{\alpha{\rm (c)}}$) order, as shown in Fig.~\ref{fig: Ca5Ir3O12}(b). 
Thus, a coexisting state with $G_z^{\rm (c)}$ and $T^{\rm (c)}_{u}$ (or $M_z^{\alpha{\rm (c)}}$) corresponds to the LT phase in Ca$_5$Ir$_3$O$_{12}$. 
It is noted that the triangles with $\pm \langle G_z^{\rm (c)} \rangle$ also have nonzero $\langle T^{\rm (c)}_{u} \rangle$ via the interplaquette coupling, and hence, the induced magnetization should be small.   

Finally, let us comment on how to distinguish $T^{\rm (c)}_{u}$ and $M_z^{\alpha{\rm (c)}}$, either of which can be the primary order parameter in the LT phase. 
Since there is correspondence as $T_u \leftrightarrow x z s_y -y z s_x$ and $M^\alpha_z \leftrightarrow (3z^2-r^2) s_z -z (x s_x + y s_y)$ based on the augmented multipole description~\cite{Hayami_PhysRevB.98.165110,kusunose2022generalization}, the quantitative difference between them appears when an external magnetic field is applied; $yz$($zx$)-type distortion is dominantly observed under the $x$-directional magnetic field in the $T^{\rm (c)}_{u}$ ($M_z^{\alpha{\rm (c)}}$) order.

To summarize, we have proposed a possible scenario for two hidden phases in Ca$_5$Ir$_3$O$_{12}$ on the basis of the augmented multipole description and the SRMS measurement. 
By considering the trimer structures consisting of EQ and MO in a multi-orbital $d$-electron model, we showed that the IT and LT phases correspond to the cluster ETD and the coexistence of the cluster ETD and MTQ (or MO), respectively. 
In the LT phase, the uniform magnetization is spontaneously induced as the secondary order parameter. 
The present results will be useful to identify hidden order parameters in Ca$_5$Ir$_3$O$_{12}$ by not only macroscopic physical responses but also local probes.

\begin{acknowledgments}
S.H. would like to thank H. Amitsuka, F. Kon, T. Hasegawa, K. Nakamura, and Y. Yamaji for their valuable discussions. 
This research was supported by JSPS KAKENHI Grants Numbers JP18H04327 (J-Physics),  JP19H04408, JP21H01037, JP22H04468, JP22H00101, JP22H01183, and by JST PRESTO (JPMJPR20L8).
Parts of the numerical calculations were performed in the supercomputing systems in ISSP, the University of Tokyo.
${}^{193}$Ir SR-based M\"{o}ssbauer spectroscopy was performed under approval of JASRI (Proposal No. 2022A1462). 
\end{acknowledgments}

\bibliographystyle{JPSJ}
\bibliography{ref}

\begin{thebibliography}{100}
\bibitem{comment_EH}
When the other two orbital basis consisting of $\ket{\pm 2
  \sigma}=\left(\ket{d_{v} \sigma} \pm i\ket{d_{xy}\sigma}\right)/\sqrt{2}$ is
  considered instead of $\ket{\pm 1 \sigma}$, $Q_v$ and $Q_{xy}$ are replaced
  by the electric hexadecapoles $Q^{\beta 1}_{4u} \propto (x^4-6x^2y^2 + y^4) $
  and $Q^{\beta 1}_{4v}\propto 4xy (x^2-y^2)$. In other words, the
  electric-hexadecapole trimer is formed. Similarly, the
  magnetic-dotriacontapole trimer consisting of $M'^{\beta 1}_{5u}\propto R_z
  (R_x^4-6R_x^2R_y^2+R_y^4)$ and $M'^{\beta 1}_{5v} \propto R_x R_y R_z (R_x^2
  - R_y^2)$ is formed instead of MO trimer.
\bibitem{comment_spring8_2}
Si(4 4 0) crystals were chosen in the present work instead of the Si(2 2 0)
  shown in Ref. 98 in order to avoid saturation of the APD detector.
\end{thebibliography}

\end{document}